\documentclass[fleqn,usenatbib]{mnras}
\usepackage{mathrsfs}
\usepackage{upgreek}

\usepackage[T1]{fontenc}
\usepackage{ae,aecompl}

\usepackage{amsmath}	
\usepackage{graphicx}
\usepackage{natbib}
\usepackage{url}
\usepackage{color}
\usepackage{amssymb}	
\usepackage{comment}

\title{The 21-cm signals from ultracompact minihaloes as a probe of primordial small-scale fluctuations}

\author[Kunihiko Furugori et al.]
{Kunihiko Furugori$^{1}$,
Katsuya T.Abe, 
Toshiyuki Tanaka, 
Daiki Hashimoto, 
\newauthor 
Hiroyuki Tashiro 
and Kenji Hasegawa
\\
$^{1}$Division of Particle and Astrophysical Sciences, Graduate School of Science, Nagoya University, Furocho Chikusa, Nagoya, 464-8602, Aichi, Japan
}

\begin{document}
\maketitle

\begin{abstract}
Ultracompact minihaloes~(UCMHs) can form after the epoch of matter-radiation equality, if the density fluctuations of dark matter have significantly large amplitude on small scales. The constraint on the UCMH abundance allows us to access such small-scale fluctuations.
In this paper, we present that, 
through the measurement of 21-cm fluctuations 
before the epoch of reionization,
we can obtain a constraint on the UCMH abundance.
We calculate the 21-cm signal from UCMHs and show that UCMHs provide the enhancement of the 21-cm fluctuations.
We also investigate the constraint 
on the UCMH abundance and small-scale curvature perturbations.
Our results indicate that the upcoming 21-cm observation, the Square Kilometre Array (SKA), 
provides the constraint on amplitude of primordial curvature power spectrum, ${\cal A}_{\zeta} \lesssim 10^{-6}$ on $100 \lesssim k \lesssim 1000~{\rm Mpc}^{-1}$.  
Although it is not stronger than the one from the non-detection of gamma-rays induced by dark matter annihilation in UCMHs, 
the constraint by the SKA will be important because this constraint is independent of the dark matter particle model. 

\end{abstract}




\section{Introduction}

Recent developments of cosmological observations allow us to access the detailed nature of the seeds of galaxies, galaxy clusters and large-scale structures.
The precise measurement of the anisotropy in the cosmological microwave background~(CMB) radiation strongly shows that the statistical property of the seeds can be expressed in the almost scale-invariant power spectrum of the curvature perturbations with the amplitude ${\cal A}_\zeta \sim 10^{-9}$~\citep{2018arXiv180706211P}, which is consistent with the prediction in the inflation paradigm~\citep{1982PhRvL..49.1110G,1983PhRvD..28..679B}.
The observations of Lyman-alpha forest
support that this tendency is confirmed up to the wave number~$k \sim 1
{\rm Mpc^{-1}}$~\citep{2006ApJS..163...80M,2011MNRAS.413.1717B}.
However, probing the perturbations below Mpc scales is still a challenge.
Currently we have the upper limit on the amplitude of the small-scale
amplitude through the constraint on the CMB distortions~\citep{2012ApJ...758...76C,2012PhRvD..86b3514D}, the primordial
black hole abundance~(\citealt{2009PhRvD..79j3520J} and see references therein) and gravitational waves, though these constraints are weaker than those on large scales~(see~\citealt{2018JCAP...01..007E} for review).

Now ultracompact minihaloes~(UCMHs) draw attention to access the perturbations on such small scales.
UCMHs are predicted to form with the excess power on small scales~\citep{2009ApJ...707..979R}.
The dark matter density fluctuations can grow after entering the horizon
even before the epoch of radiation-matter equality.
Therefore, the overdensity regions with the density fraction~$\delta >10^{-3}$
at the horizon entry can collapse to minihaloes around $z \sim 1000$
well before the standard structure formation history.
Since UCMHs could be formed by the radial infalling in high redshifts,
UCMHs are assumed to have 
a more compact profile
with a larger central density than 
typical dark matter halos 
in the standard hierarchical structure formation~\citep{2009ApJ...707..979R}.
The first cosmological simulation
of the UCMH formation has been done by
\citet{2017PhRvD..96l3519G}.
They showed that
a large amplitude on small scales
leads to the early structure formation
and the resultant dark matter halos have 
a large density.
Recent numerical simulations by~\citet{2018PhRvD..97d1303D,Delos+:2018a} showed that UCMHs originated from
the spike-shape spectrum on small scales
have the Moore profile, $\rho \propto r^{-3/2}$ at the inner cusp region~\citep{1999MNRAS.310.1147M}, which is steeper than the Navarro-Frenk-White~(NFW) profile~\citep{10.1093/mnras/275.3.720}.

If the dark matter is a weakly interacting massive particle~(WIMP)~\citep{1985NuPhB.253..375S,1996PhR...267..195J,1999dmap.conf..592K}, 
then UCMHs can become cosmological gamma-ray sources 
because the signal of the annihilation of WIMPs is enhanced in the dense region at the centre of UCMHs.
Currently, the non-detection of the gamma-ray emission from the WIMP annihilation provides the constraint on the UCMH abundance and the stringent constraints on the small-scale curvature power spectrum,
$A_{\zeta} < 10^{-7}$ for $10 <k < 10^8~{\rm
Mpc}^{-1}$~\citep{2010PhRvD..82h3527J, 2010PhRvL.105k9902S, 2012PhRvD..85l5027B,
2016MNRAS.456.1394C,2018PhRvD..97b3539N,2018PhRvD..97d1303D,Delos+:2018a}.
The non-detection of neutrino from the WIMPs also provides a similar
constraint~\citep{2013PhRvD..87j3525Y,2018PhRvD..97b3539N}. 
The Thomson scattering optical depth for the CMB photons also provides the UCMH abundance because the annihilation gamma-ray from UCMHs can contribute
to the photon budget for cosmic reionization~\citep{2011MNRAS.418.1850Z,2011JCAP...12..020Y,2016EPJP..131..432Y,2017JCAP...05..048C}. Observations of gravitational lensing by UCMHs also constraint the UCMH abundance   ~\citep{2012PhRvD..86d3519L,2016MNRAS.456.1394C}.

The aim of this paper is to demonstrate that
future 21-cm observations can provide the constraint on the UCMH abundance and small-scale curvature power spectrum, which does not require the dark matter particle model. The measurements of 21-cm line emitted by hyperfine transition of neutral hydrogen form high redshifts universe are expected to be a powerful tool to probe the structure formation during the dark ages to the epoch of reionization~(EoR)~\citep{2006PhR...433..181F}. The signal amplitude strongly relates to the spatial distribution of neutral hydrogen in the intergalactic medium~(IGM) and collapsed objects. 
The 21-cm observations can probe the IGM matter fluctuations on much smaller scales than the Silk scales~\citep{2004PhRvL..92u1301L}.
Minihaloes which formed in the early stage of the hierarchical structure formation history, are also promised source of 21-cm signal in high  redshifts~\citep{2002ApJ...572L.123I,2006ApJ...652..849F}.
Recent studies have shown that the measurements of their signals by future observations provide the detailed stochastic nature of small-scale density
fluctuations including the running spectrum, non-Gaussianity and so on~\citep{2012MNRAS.426L..21C,2018JCAP...02..053S,2019JCAP...02..033S}.
It was also suggested that 21-cm observations can provide the limit on the Primordial Black Hole~(PBH) abundance indirectly~\citep{2008arXiv0805.1531M,2013MNRAS.435.3001T}.
Nowadays, several projects are conducting observation to detect the
21-cm fluctuations around and before the EoR, e.g.,~the Low Frequency Array \citep{2013A&A...556A...2V},
the Giant Meterwave Radio Telescope
\citep{2011MNRAS.413.1174P}, the Murchison Widefield Array \citep{2013PASA...30....7T,2013PASA...30...31B}and the Precision Array for probing
the EoR
\citep{2010AJ....139.1468P}.
Since the measurement of 21-cm fluctuations through these observations is very difficult, there are only the upper limit on the 21-cm　fluctuations~\citep{2017ApJ...838...65P,2013MNRAS.433..639P,2015ApJ...809...61A,2016ApJ...833..102B,2019ApJ...887..141L}.
However, it is expected that the high sensitivity of the future instrument, the Square Kilometre Array~(SKA), can measure the 21-cm fluctuations up to the redshift~$z\sim 28$ ~\citep{2015aska.confE...1K}.

In this paper, we discuss the possibility of SKA to measure the 21-cm fluctuations originated in UCMHs. Constructing the baryon gas model in UCMHs, we first evaluate the 21-cm fluctuations created by UCMHs. Then we evaluate the detectability of these 21-cm fluctuations by the SKA and discuss the possible constraint on both of the UCMH abundance and the small-scale primordial curvature perturbations.

This paper is organised as follows. In section~\ref{sec:ucmh_review}, we present a brief review about UCMHs. In section \ref{sec:theory}, we construct the baryon gas model in UCMHs considering the hydrostatic equilibrium with the compact dark matter profile in UCMHs. 
Then we evaluate the 21-cm signal from an individual UCMH. In section~\ref{flusec}, we calculate the 21-cm fluctuations due to the UCMHs distribution. 
We also discuss the detectability of these fluctuations by the SKA. 
Based on the discussion,
we provide the possible constraint on the UCMHs and the primordial curvature perturbations. 
Finally we give a conclusion of this paper in section~\ref{conc}.

Throughout this paper we assume the flat $\Lambda \rm{CDM}$ cosmology with Hubble constant
$h=0.701$, matter density $\Omega_{\rm{m}}=0.1408 h^{-2} $, baryon
density $\Omega_{\rm{b}}=0.022 h^{-2} $, dark matter density
$\Omega_{\rm{dm}}=0.1187 h^{-2} $, scale invariant spectral index
$n_{\rm{s}}=0.965$, and power spectrum normalized factor $\sigma_{8}=0.8$.

\section{UCMHs with the spiked matter spectrum}
\label{sec:ucmh_review}
Small-scale matter density fluctuations with larger amplitude can be seeds of UCMHs.
In this section, we begin with the brief review of UCMHs with the spiky shape power spectrum, according to~\citet{2018PhRvD..97d1303D,Delos+:2018a}.

Adding on to the nearly scale-invariant spectrum,
we consider the spike shape of the primordial curvature power spectrum on a small scale, which enters the horizon during the radiation dominated era,
\begin{equation}
\mathcal{P}_{\zeta}(k)=\mathcal{A}_{\zeta} k_{\rm s} \delta(k-k_{\rm s}), \label{curvature}
\end{equation}
where $\mathcal{A}_{\zeta}$ is the amplitude of the spike shape and $k_{\rm s}$ is the wave number of the spike.
During the radiation dominated era, the dark matter fluctuations can grow logarithmically.
Here, for simplicity, we adopt the Dirac delta function to represent the additional spike-shape spectrum. 
Then after the radiation-matter equality, they evolve proportionally to the scale factor. 
When the density amplitude in a overdensity region reaches the critical value for the collapse, the overdensity region can collapse to a dark matter halo as the standard hierarchical structure formation.

However, when the spike amplitude is much larger than one of the ordinary scale-invariant spectrum,
the properties of the resultant dark matter halo are different from that of the standard dark matter halos. 
N-body simulations by \citet{Delos+:2018a} showed that the spike spectrum produces the isolated distribution of compact dark matter
halos, that is UCMHs. 
The resultant dark matter density profile in a UCMH is represented by a self-similar form as
\begin{equation}
\rho_{\rm dm}(r)=\rho_{\rm s}y_{\rm dm}\left(\frac{r}{r_{\rm s}}\right),
\label{ydm}
\end{equation}
where $r$ is the radial distance from the centre, $\rho_{\rm s}$ is the scale density
and $r_{s}$ are the scale radius. The non-dimensional density profile~$y_{\rm dm}$ is given by
\begin{equation}
y_{\rm dm}(x)=\frac{1}{x^{\alpha} (1+x)^{3-\alpha}},
\end{equation}
with defining $x \equiv r/r_{\rm s}$.
The simulations demonstrated that
the index $\alpha$ for UCMHs
is $\alpha = 1.5$, that is the Moore profile~\citep{1999MNRAS.310.1147M}.
It is known that the hierarchical structure formation yields the NFW profile~\citep{10.1093/mnras/275.3.720} whose index corresponds to $\alpha =1$. Therefore, UCMHs have a steeper dark matter profile in the inner region.

\citet{Delos+:2018a}~also showed that
$\rho_{s}$ and $r_{s}$ are related to
the UCMH forming redshift $z_{\rm{c}}$ and the wave number of the spike as
\begin{align}
\rho_{\rm s} &= 30(1+z_{\rm{c}})^3 \Omega_{\rm m} \rho_{\rm crit, 0} \label{eq:rhos}\\
r_{\rm s} &= 0.7[(1+z_{\rm{c}})k_{\rm s}]^{-1}, \label{eq:rs}
\end{align}
where $\rho_{\rm crit, 0}$ is the critical density at present.

For the UCMH~(virial) mass,~$M_{\rm vir}$,
we adopt the mass enclosed within $r_{200}$,
\begin{align}
M_{\rm vir} &=
4\pi \rho_{\rm s} r_{\rm s}^{3} m(u_{\nu}),~\label{1}\\
m(x) &\equiv \int^{x}_{0} u^2 y_{\rm dm}(u){\rm d}u.
\end{align}
where $u_{v}$ is $u_{v} \equiv r_{200}/r_{\rm s}$
and  the scale $r_{200}$ is defined as the scale inside which the averaged dark matter density is 200 times the mean dark matter density of the Universe. We found that the UCMH mass is related to $z_{\rm c}$ and $k_{\rm s}$ as in
\begin{equation}
M_{\rm  vir} \sim 4 \times 10^{3}~{\rm M_{\odot}} \times \left(\frac{k_{\rm s}}{10^{3} ~ {\rm Mpc}^{-1}} \right)^{-3} \ln  \left( \frac{1+z_{\rm c}}{1+z} \right). \label{eq:vir_ks_z}
\end{equation}

Therefore, the virial mass of a UCMH grows logarithmically even after the formation of the UCMH.
This growth corresponds to the late-time accretion on the outer region suggested
in the simulation~\citep{Delos+:2018a}.
The number density of UCMHs can be evaluated from the peak
theory~\citep{1986ApJ...304...15B}, because UCMHs form at the peak locations
of the density fluctuations following the peak shape spectrum.
The differential UCMH number density by the formation redshift~$z_{\rm c}$
can be expressed in
\begin{equation}
\frac{dn}{dz_{\rm c}} =\frac{k_{\rm s}^3}{1+z_{\rm c}}
h\left[\frac{\delta_{\rm c}}{\sigma_{\rm d}(z_{\rm c})}\right],
\label{eq:dndz}
\end{equation}
where
$\delta_{\rm c}$ is the critical density contrast for collapse,~$\delta_{\rm c}=1.68$. $\sigma_{\rm d}(z)$ is the root-mean-squared~(rms) variance
of the density fluctuation given by the peak shape spectrum at the
redshift~$z$.
In equation~(\ref{eq:dndz}), $h(\nu)$ is given by
\begin{equation}
h(\nu)=\frac{\nu}{(2\pi)^2 3^{3/2}}e^{-\nu^2/2}f(\nu).
\label{hnu}
\end{equation}
where $f(\nu)$ is provided in
equation~(A15) of~\citet{1986ApJ...304...15B}.
For the detailed derivation of equation~\eqref{eq:dndz},
we refer readers to the appendix in~\citet{Delos+:2018a}.
Through equation~\eqref{eq:dndz}, the abundance of UCMHs is related to
the spike shape properties,~${\cal A}_\zeta$ and $k_{\rm s}$.
Hence, once we obtain the constraint on the abundance of UCMHs,
we can provide the limit on the peak-shape spectrum through equation~\eqref{eq:dndz}.
\section{twenty one cm signal from a single UCMH}
\label{sec:theory}

In this section, we evaluate the 21-cm signal from a single UCMH.
The 21-cm signal is sensitive to the gas density profile in a UCMH. 
Therefore, we first derive the profile in a UCMH assuming the hydrostatic equilibrium with the gravitational potential of the dark matter profile presented in the previous section.

\subsection{Baryon gas mass in UCMHs}

Baryonic density fluctuations, unlike the dark matter,
cannot grow before the decoupling of photons.
After the decoupling, the baryon density fluctuations start to evolve,
following the dark matter density fluctuations.
However, since the baryon gas resists the gravitational collapse, there is the critical scale for the collapse, the so-called Jeans scale.
Therefore, to evaluate the baryon gas mass inside UCMHs, we consider two scenarios
depending on the spike scale,~$k_{\rm s}$.

When $k_{\rm s}$ is smaller than the Jeans wave number,~$k_{\rm J}$,
the baryon density fluctuations with $k_{\rm s}$ can grow and collapse,
following the dark matter density evolution.
Therefore, for simplicity, we assume that UCMHs can have the baryon gas whose mass ratio to dark matter is the same as the ratio of the cosmological background,
\begin{equation}
 M_{\rm gas}(z) = \frac{\Omega_{\rm b}}{\Omega_{\rm dm}} M_{\rm vir}(z)~\quad {({\rm
  for}~k_{\rm
  s } < k_{\rm J})}.
  \label{baryacc1}
\end{equation}

On the other hand,
when $k_{\rm s}$ is larger than $k_{\rm J}$,
the baryon density fluctuations with $k_{\rm s}$ cannot evolve due to
its own pressure.
However,
UCMHs can host dense baryon gas through the accretion of baryon gas.
We obtain the accreted baryon gas mass at the redshift $z$ from
\begin{equation}
M_{\rm{gas}}(z)=
\int^{z_{\rm{acc}}}_{z}
dz'\frac{\dot M_{\rm gas}(z')}
{ (1+z')H(z')}~\quad {({\rm
  for}~k_{\rm
  s } > k_{\rm J})},
\label{baryacc2}
\end{equation}
where
$z_{\rm acc}$ is the starting redshift of the gas accretion.
We set $z_{\rm acc}$ to $z_{\rm acc}={\min}[z_{\rm c}, ~z_{\rm dec}]$ where $z_{\rm dec}$ is the redshift for the decoupling of photons.
For the accretion rate,~$\dot M_{\rm gas}(z)$,
we adopt the Bondi-Hoyle-Lyttleton accretion~\citep{1944MNRAS.104..273B},
\begin{equation}
\dot{M}_{\rm{gas}}(z)={4 \pi G^2 \Omega_{\rm{b}} \rho_{\rm{crit}}(z) M_{\rm{vir}}^{2}(z)}{v_{\rm{r}}^{-3}(z)}.
\label{bondi}
\end{equation}
Here $v_{\rm r}$ is the relative velocity between baryon and dark matter, $v_{\rm r}(z)=30~{\rm km s^{-1}}[(1+z)/1000]$~\citep{2010PhRvD..82h3520T}\footnote{Here, we neglect the contribution of
the baryon thermal velocity, which is smaller than the relative velocity between
baryon and dark matter~\citep{2010PhRvD..82h3520T}. The peak spectrum
also causes the additional relative velocity. However, the velocity is
suppressed by the wave number~$k$. Therefore, we also neglect this
contribution.}. Note that, to avoid the over-accretion to UCMHs,
we set the upper limit of the accretion gas mass, $M_{\rm gas} <
M_{\rm vir}{\Omega_{\rm b}}/{\Omega_{\rm dm}}$.

\begin{figure}
 	\includegraphics[width=\columnwidth]{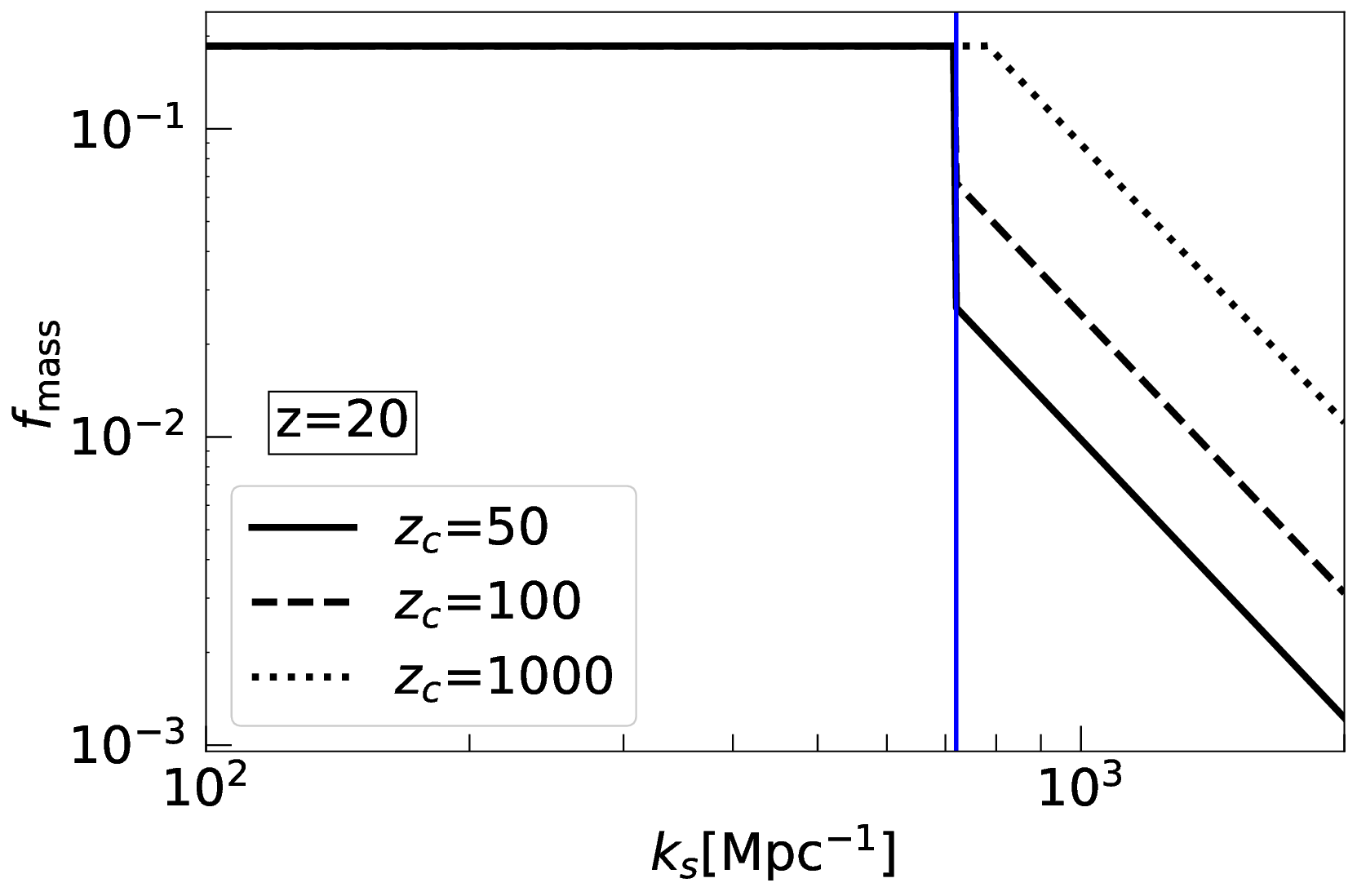}
     \caption{Baryon gas mass ratio to the dark matter mass,~$f_{\rm{mass}}=M_{\rm{gas}}/M_{\rm{vir}}$, at the redshift~$z=20$. The solid, dashed, and dotted black lines show the baryon gas mass fraction of the UCMH formed at $z_{\rm c} =50,~100$, and $1000$, respectively. We also provide the Jeans wave number at $z =20$ in a vertical blue line.}
     \label{graph1}
 \end{figure}

In figure~\ref{graph1},
we plot the baryon gas mass ratio to the dark matter mass in UCMHs,~$f_{\rm{mass}}=M_{\rm{gas}}/M_{\rm{vir}}$, 
at the redshift~$z=20$ as a function of the spike wave number~$k_{\rm{s}}$.
The black solid, dashed and dotted lines represent $f_{\rm mass}$ for UCMHs collapsed at $z_{\rm c}=50$,~$100$,~and~$1000$, respectively. For reference, we plot the Jeans wave number at $z=20$ in the blue thin vertical line. Through this paper, we adopt the Jeans wave number in \citep{2001PhR...349..125B}.

When the spike wave number is smaller than $k_{\rm J}$ we assume that the baryon can collapse following the dark matter. 
Therefore, the gas mass ratio is $f_{\rm{mass}}= {\Omega_{\rm b}}/{\Omega_{\rm dm}}$ as shown in equation~\eqref{baryacc1}. 
On the other hand, when $k_{\rm s} > k_{\rm J}$, UCMHs obtain the baryon gas through the accretion with the accretion rate in equation~\eqref{bondi}.
Since the accretion rate is proportional to $M_{\rm{vir}}^2 \propto k_{\rm s}^{-6}$, the baryon gas mass steeply declines with increasing $k_{\rm s}$. Figure~\ref{graph1} also shows that a UCMH with large $z_{\rm c}$ has high gas mass ratio. When the UCMH formation happens in higher redshifts, the UCMH undergoes the long duration of the accretion. As a result, the earlier the UCMH forms, the larger the accreted gas mass is in the UCMH.

\subsection{Baryon gas density and temperature in UCMHs}
\label{sec:gaspro}

Now we evaluate the density and temperature profiles of the baryon gas following the method of~\cite{2001MNRAS.327.1353K}, which is based on the hydrostatic equilibrium assumption.

Since the dark matter profile is expressed in the self-similar form, the gas density profile of UCMHs, $\rho_{\rm gas}$, would also be the self-similar form as
\begin{equation}
\rho_{\rm{gas}}(r)=\rho_{\rm{gas}}(0) y_{\rm{gas}}\left(\frac{r}{r_{\rm s}}\right), \label{2-2}
\end{equation}
where $y_{\rm gas}(x)$ is the non-dimensional gas profile normalized as $y_{\rm{gas}}(0)=1$.
Applying the polytropic gas model with the polytropic index~$\gamma$, we can write the gas temperature profile in
\begin{equation}
T_{\rm{gas}}(r)=T_{\rm gas}(0)y_{\rm gas}^{\gamma-1}\left(\frac{r}{r_{\rm s}}\right). \label{3}
\end{equation}

The assumption of the hydrostatic equilibrium allows us to relate the gas pressure profile to the dark matter density profile,
\begin{equation}
{\rho^{-1}_{\rm gas}} \frac{dP_{\rm gas}}{dr}=-\frac{GM(r)}{r^2}. \label{4}
\end{equation}
Here we assume that the gas component does not contribution to the gravitational potential. 
To derive the pressure profile, we adopt the equation of state of ideal gas, 
\begin{equation}
P_{\rm gas}=\frac{k_{\rm{B}} T_{\rm gas}}{\mu m_{\rm{p}}}\rho_{\rm gas}  \label{5},
\end{equation}
where $k_{\rm B}$ is the Boltzmann constant, $m_{\rm{p}}$ is the proton mass and $\mu$ is the mean molecular weight of the gas. 
Using equations~\eqref{3} and \eqref{4}, $y_{\rm gas}$ can be derived as~\citep{1998ApJ...509..544S}
\begin{equation}
y^{\gamma-1}_{\rm gas}(x)=1- 3\eta^{-1}_0 \frac{\gamma-1}{\gamma} \frac{u_{v}}{m(u_{v})} \int^{x}_{0} du \frac{m(u)}{u^2}\label{eq:ygas},
\end{equation}
where the mass-temperature normalization factor $\eta_0$ is expressed as
\begin{equation}
\eta_0=\frac{3k_{\rm B} r_{\rm{200}} T_{\rm gas}(0)} {G\mu m_{\rm p}  M_{\rm{vir}}}.
\end{equation}

To obtain the gas profile, it is required to fix $\eta_0$ and $\gamma$. 
For the determination of these parameters, we take the assumption that the gas profile traces the dark matter profile outside halo core. This condition is satisfied by imposing the slopes of these two profiles to match,
\begin{equation}
s_{*}\equiv \left.\frac{\rm d\ln \rho_{\rm{dm}}(x)}{\rm d\ln x}\right|_{x=x_{*}}=\left.\frac{\rm d\ln \rho_{\rm{gas}}(x)}{\rm d\ln x}\right|_{x=x_{*}}, \label{eq:sd}
\end{equation}
where $x_{*}$ is the location outside the core region. As a result, we obtain
\begin{align}
\gamma&=1 -
\frac{1}{s_{*}} +\frac{ \partial \ln[m(x_{*})/s_{*}]}{s_{*} \partial \ln x_{*}}, \label{gamma0}
\\
\eta_0&=3\gamma^{-1}\left[
  \left(\frac{-1}{s_{*}}\right)
    \left[\frac{x^{-1}_{*}m(x_{*})}{u_{v}^{-1}m(u_{v})}\right]+
  ({\gamma-1}) \frac{u_{v}}{m(u_{v})}
    \int^{x_{*}}_{0} du \frac{m(u)}{u^2}\right], \label{eta0f}
\end{align}
where $s_{*}$ is provided from the dark matter distribution,
\begin{equation}
s_{*}=-\left[\alpha+(3-\alpha) \frac{x_{*}}{1+{x_{*}}}\right].
\end{equation}

One can see that $\eta_0$ is a function of $x_*$ and $\gamma$. It is preferable that $\eta_0$ does not depend on the location $x_*$. To satisfy this condition, we impose the following condition, according to \cite{Komatsu+:2002},
\begin{equation}
\frac{\partial \eta_0}{\partial x_*} =0,\label{eq:sd2}
\end{equation}
with setting $x_*=u_{v}$.
This equation yields
\begin{equation}
\gamma=\frac{16u_{v}^2+20u_{v}+5}{3(1+2u_{v})^2}-\frac{2u_{v}}{3(1+2u_{v})m(u_{v})} \left(\frac{u_{v}}{1+u_{v}}\right)^{1/2} \label{gaM},
\end{equation}
where $m(x)$ is given by
\begin{equation}
m(x)=2 \ln(\sqrt{x}+\sqrt{1+x}) -2 \sqrt{\frac{x}{1+x}}.
\end{equation}

The final parameter to determine the gas profile is
$\rho_{\rm gas}(0)$, which we obtain through
\begin{equation}
\rho_{\rm{gas}}(0)={M_{\rm{gas}}}
\left[
{4 \pi r_{\rm s}^{3} \int^{u_{v}}_{0} y_{\rm{gas}}(u)u^2\mathrm{d}u}
\right]^{-1},
 \label{rhogas}
\end{equation}
where $M_{\rm gas}$ is given in equations~\eqref{baryacc1}
and~\eqref{baryacc2}.

\subsection{Brightness temperature} \label{bris}

Let us evaluate
the 21-cm signal of a single UCMH at $z$.
First we consider a line of sight intersecting the UCMH
with the impact parameter~$\alpha_{\rm R}$~(in unit of $r_{200}$) from
its centre.
The brightness temperature along this line of sight is
given by
\begin{equation}
T_{\rm{b}}(\alpha_{\rm R})=T_{\rm{CMB}}(z)e^{-\tau(\alpha_{\rm R})}+\int ^{R_{\rm
 max}}_{-R_{\rm max}} T_{\rm{s}}(l) e^{-\tau(\alpha_{\rm R},R)} \frac{\mathrm{d} \tau}{\mathrm{d} R}\mathrm{d}R,
\label{eq:Tb_alpha}
\end{equation}
where
$T_{\rm CMB}(z) = T_{\rm CMB,0} (1+z)$
with $T_{\rm CMB,0} = 2.73\, \rm{K}$,
$R$ is the coordinate along the line of sight, 
whose origin is set to the centre of the UCMH,
$T_{\rm s}$ is the radial profile of the spin temperature in the
UCMH~(we will discuss later), 
$l$ represents the radial distance satisfying $l^2=R^2+(\alpha_{\rm R}
r_{200})^2$
and $R_{\rm max}$ is defined as $R_{\rm max}^2 \equiv {r_{200}^{2}(1-\alpha_{\rm R}^2)}$.

The optical depth, $\tau (R)$, 
for the frequency at the rest frame of the UCMH, $\nu$, is calculated from
\begin{equation}
\tau (\alpha_{\rm R},R)=\frac{3c^2 A_{10} T_{*}}{32 \pi \nu_{*}^2}
 \int^{\rm R}_{-R_{\rm max}} \frac{n_{\rm Hi}(l)\phi(\nu,l)}{T_{\rm s}(l)} \mathrm{d}R,
\label{eq:tau_R}
\end{equation}
where $T_*$ is $T_* =0.0681~\rm K$ and $\nu_* = 1440$~MHz,
$A_{10}$ represents the Einstein A coefficient for the 21-cm transition,
$A_{10}=2.85
\times 10^{-15} {\rm s}^{-1}$, and $n_{\rm Hi}(r)$ provides the radial profile
of neutral hydrogen which we obtain through $n_{\rm Hi}(r) = (1-Y)\rho_{\rm
gas}/m_{\rm p} $ with the helium fraction~$Y$.
Here $\phi(\nu,r)$ is the line profile at the radial distance~$r$ for the rest-frame frequency~$\nu$, which suffers the Doppler
broadening due to the thermal velocity of the gas,
\begin{equation}
\phi(\nu , r) = \frac{1}{\Delta \nu \sqrt{\pi}} \exp \left(-\frac{(\nu - \nu_{*})^2}{\Delta \nu^2} \right),
\end{equation}
where the Doppler width $\Delta \nu$ is given by
\begin{equation}
\Delta \nu=\frac{\nu_*}{c} \sqrt{ \frac{2 k_{\rm B} T_{\rm{gas}}(r)}{m_{\rm H}}}.
\end{equation}
In equation~\eqref{eq:Tb_alpha},
$\tau (\alpha_{R})$ is obtained by $\tau (\alpha_{R})=  \tau
(\alpha_{R},R_{\rm max})$.
Hereafter, we set $\nu$ to $\nu=\nu_*$~[that is, the observed frequency is $\nu_{\rm obs} =(1+z) \nu_*$].
We provide a comment about this setting in section~4.

To calculate equation~\eqref{eq:Tb_alpha} with equation~\eqref{eq:tau_R},
it is required to evaluate the spin temperature which is related to the
ratio of the number densities of the two hyperfine structure levels.
The spin temperature is given by~\citep{1959ApJ...129..536F},
\begin{equation}
T_{\rm s}^{-1}=\frac{T_{\rm{CMB}}^{-1}(z)+x_{\rm c} T_{\rm{gas}}^{-1}+x_{\alpha} T_{\alpha}^{-1}}{1+x_{\rm c}+x_{\alpha}}, \label{ts}
\end{equation}
where
$T_{\alpha}$ is the colour temperature of Ly$\alpha$ photons, and
$x_{\rm{c}}$ and $x_{\alpha}$ represent the coupling coefficients for
the gas collisions and the Ly$\alpha$ pumping, respectively.
In this paper, we set $x_{\alpha} = 0$, because we assume that there
exist no UV and X-ray external sources.
The coupling coefficient for gas collisions is expressed in
\begin{equation}
x_{\rm c}=\frac{T_{*}}{A_{10} T_{\rm{CMB}}}C_{\rm{Hi}},
\end{equation}
where $C_{\rm Hi} $ is the collisional coefficient between Hi atoms and we adopt the value in \cite{2006ApJ...637L...1K}.
Note that, since we assume that the gas inside UCMHs as fully neutral one, we neglect the contribution from the collisions of Hi with protons and electrons.
Using equation~\eqref{ts} with the gas radial profile obtained in section~\ref{sec:gaspro},
we obtain the radial profile of the spin temperature in the UCMHs for
equations~\eqref{eq:Tb_alpha} and~\eqref{eq:tau_R}.

The effective brightness temperature averaged over the single UCMH
cross-section,~${\cal S}=\pi r_{\rm{200}}^2$, is given by
\begin{equation}
T_{{\rm b}, z_{\rm c}}=\frac{\int T_{\rm b} \mathrm{d} {\cal S}}{\cal S}=2 \int^{1}_{0}T_{\rm b}(\alpha_{\rm R}) \alpha_{\rm R} \mathrm{d} \alpha_{\rm R}.
\end{equation}

We measure the 
21-cm signals as 
the difference of the brightness temperature from the CMB temperature,
which is called the differential brightness temperature.
The differential brightness temperature for the single UCMH at the
redshift~$z$ can be 
calculated from
\begin{equation}
\delta T_{{\rm b}, z_{\rm c}}(z)=\frac{T_{{\rm b} , z_{\rm c} }}{1+z}-T_{\rm CMB}(0).
\end{equation}

\begin{figure}
 	\includegraphics[width=\columnwidth]{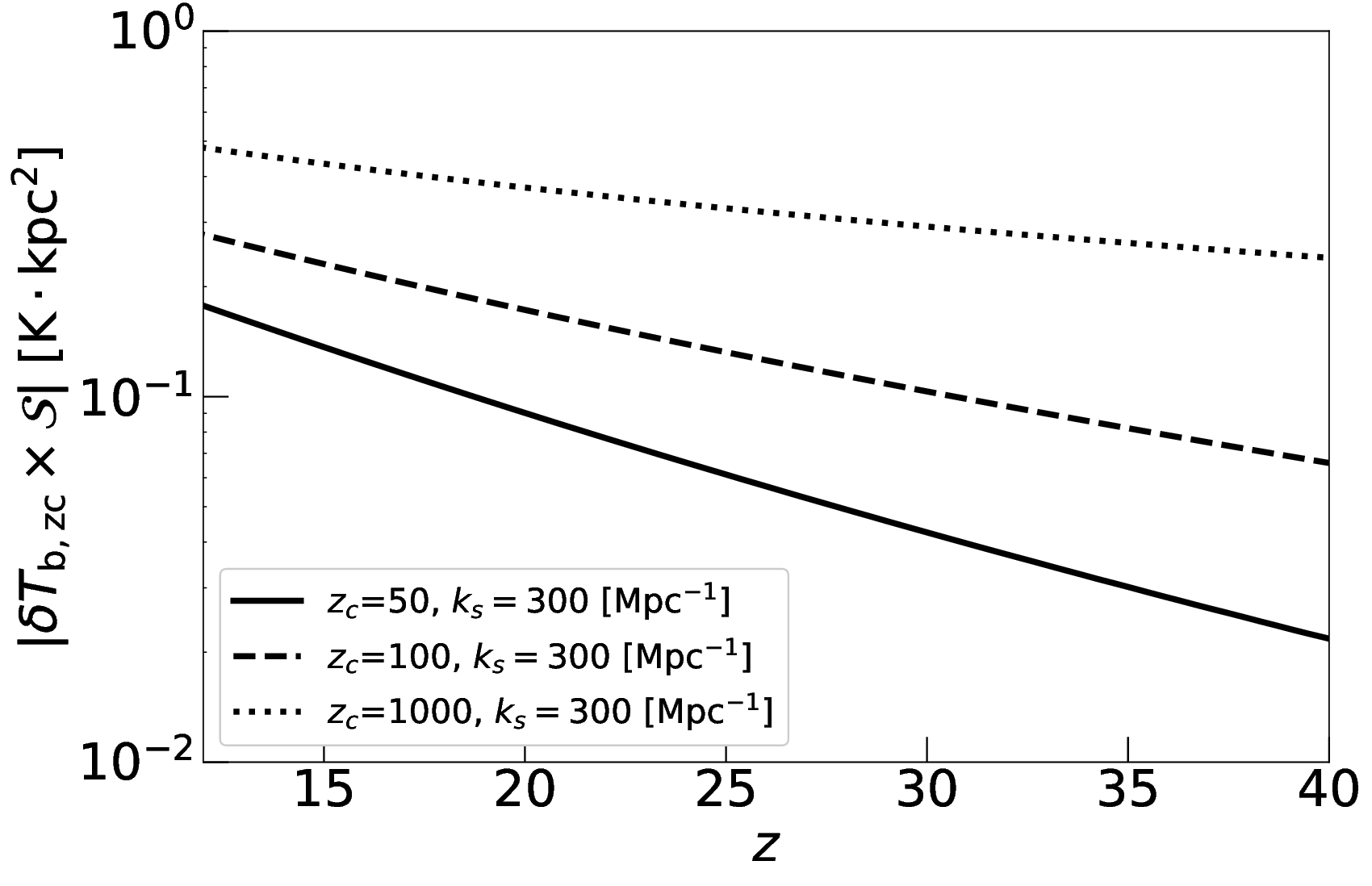}
     \caption{The 21-cm signal from a single UCMH, which is given by a product of the differential brightness temperature and a geometrical cross-section of UCMH, as a function of redshift. We show the 21-cm signal from a single UCMH at the formation redshift $z_{\rm c}=50,100,1000$ as the solid, dashed and dotted black lines, respectively.}
     \label{fig:signle_resdhift_evo}
 \end{figure}

Figure~\ref{fig:signle_resdhift_evo} provides the evolution of the 21-cm
signal from the single UCMH.
In this figure, the $y$-axis is  
the product of the differential brightness temperature and the UCMH
cross-section, which corresponds to the total 21-cm flux from the single UCMH.
Here we set $k_{\rm s} = 300~{\rm Mpc}^{-1}$ which is smaller than $k_{\rm J}$
at $z=10$. In this case, the signal is observed as emission on the
CMB frequency spectrum.
The 21-cm signal becomes strong as the baryon gas mass or temperature in the UCMH increases. 
Therefore, the signal monotonically becomes larger as the redshift decreases, because the UCMH virial mass grows logarithmically due to the accretion as shown in equation~\eqref{eq:vir_ks_z} even with fixing~$k_{\rm s}$.

In figure~\ref{fig:signle_resdhift_evo}, we also show the dependence of the signal on the collapse redshift,~$z_{\rm c}$. 
UCMHs collapsed in higher redshifts have larger virial mass, according to equation~\eqref{eq:vir_ks_z}.
Hence, the UCMH with higher $z_{\rm c}$ provides the stronger signal, compared to the one with small $z_{\rm c}$.

\begin{figure}
 	\includegraphics[width=\columnwidth]{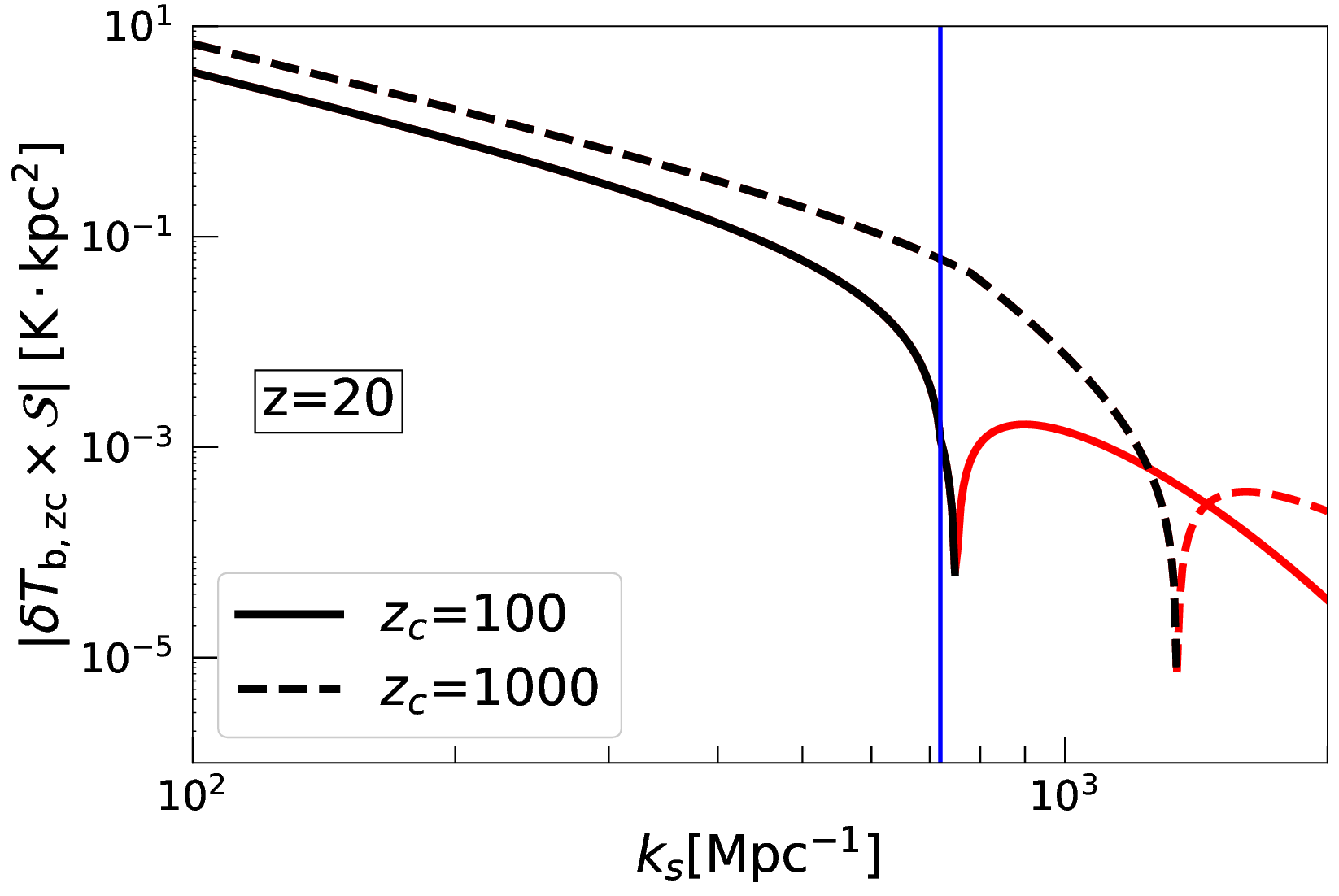}
     \caption{Dependence of the 21-cm signal from a single UCMH on the spike scale~$k_{\rm s}$. The solid and dashed lines correspond to the 21-cm signal from a single UCMH at $z_{\rm c}=100 and 1000$, respectively. The vertical line show the Jeans wave number at $z=20$.}
     \label{fig:single_ks}
 \end{figure}

We plot the signal as a function of $k_{\rm s}$ in figure~\ref{fig:single_ks}.
The black lines represent the emission signals, while the red ones
denote the absorption signal.
Larger $k_{\rm s}$ provides UCMHs with small gas mass and low baryon gas temperature.
Therefore, as $k_{\rm s}$ becomes larger, the spin temperature decreases and, 
then, the signal shifts from the emission to absorption because the averaged spin temperature cannot exceed the CMB temperature in such low UCMH mass.
The figure also tells us that the signal radically decreases when
$k_{\rm s}$ is larger than $k_{\rm J}$.
This is because the baryon with such $k_{\rm s}$ cannot collapse to the
UCMH and the baryon gas mass fraction drops down
as shown in figure~\ref{graph1}.

\section{the 21-cm fluctuations due to the UCMH clustering} \label{flusec}

The angular scale of an individual UCMH on the sky is too small to be resolved even by upcoming 21-cm observations such as the SKA. 
Therefore, the key observable for the UCMHs is the differential brightness temperature fluctuations due to the number density fluctuation of UCMHs in the observational beam.
In this section, we evaluate the rms fluctuations due to the UCMH as the 21-cm signal of UCMHs.

Now we consider the ensemble of UCMHs formed in the redshift from
$z_{\rm{c,max}}$ to $z_{\rm{c,min}}$.
Following~\citet{2002ApJ...572L.123I},
the mean 21-cm differential brightness temperature from this ensemble at the redshift~$z$ is given by
\begin{equation}
\overline{\delta T_{\rm b}}(z) = \frac{c(1+z)^4}{\nu_*
 H(z)}\int^{z_{\rm{c,max}}}_{{z_{\rm c,min}}} \Delta \nu_{\rm eff}(z)
 \delta T_{{\rm b},z_{\rm c}}(z)
{\cal  S}\frac{dn}{dz_{\rm c}} \mathrm{d}z_{\rm c}.
\label{eq:barTb}
\end{equation}
Here we set $z_{\rm c, max}=4000$ and $z_{\rm c, max}=50$, and
$\Delta \nu_{\rm{eff}}$ is 
the effective linewidth, which is provided by
\begin{equation}
\Delta \nu_{\rm{eff}}(z)=[\phi(\nu_*)(1+z)]^{-1}
 \approx\frac{\nu_*}{c(1+z)} \sqrt{\frac{2 \pi k_{\rm B} \overline{T}_{\rm{gas}}(z)}{m_{\rm{H}}}},
\end{equation}
where $\overline{T}_{\rm gas}(z)$ is the volume averaged temperature within an individual UCMH.
To calculate the mean 21-cm differential brightness temperature, it is required to evaluate the total flux integrated over the line profile.
Instead of this integration,
we take the rectangular approximation whose width is 
$\Delta \nu_{\rm eff}$ and 
height is obtained by setting $\nu = \nu_{\rm *}$ for the optical depth 
in equation~\eqref{eq:tau_R}~(for more detailed discussion, we refer readers to section 2 in~\citealt{2002ApJ...572L.123I}).

On large scales, UCMHs are distributed homogeneously with the number
density $n = \int \mathrm{d}z_{\rm c} \mathrm{d}n/ \mathrm{d}z_{\rm c}$. However, as the scale-invariant
fluctuations on large scales evolve,
the distribution of UCMHs can gravitationally trace this fluctuations on
large scales.
Therefore the number of UCMHs in an observational beam volume
is fluctuated, depending on the matter density fluctuations in the beam.
The rms fluctuations due to UCMHs can
be evaluated by 
\begin{equation}
\langle \delta T_{\rm b}^2 \rangle^{1/2}(z)
= \beta(z) \sigma_{\rm p}(z) \overline{\delta T_{b}}(z),
\label{ucmhflu}\end{equation}
where $ \sigma_{\rm p} $ is the rms fluctuation of the matter density averaged in the observation beam volume.
We adopt the cylinder-shape beam: the diameter and height of the cylinder-shape beam respectively correspond to the angular resolution, $\Delta \theta$, and the frequency resolution, $\Delta \nu$, of the observations.
In this configuration, $\sigma_{\rm p}$ is given by
\begin{equation}
\sigma_{\rm p}(z)=\int^{\infty}_{0} \frac{\mathrm{d}k}{k} \mathcal{P}(k,z) W_{\rm{cy}}(k,z),\label{obsvari}
\end{equation}
where $\mathcal{P}(k,z)$ is the non-dimensional matter power spectrum at the redshift~$z$. For the matter power spectrum, we use the amplitude
$\sigma_8 = 0.81$ and the spectral index~$n_{\rm s} =0.965$.
The cylinder-shape beam window function is expressed in
\begin{equation}
 W_{\rm{cy}}(k,z)
 =\frac{16}{R^2 L^2}
 \int^{1}_{0} \mathrm{d}x \frac{\sin^2(kLx/2)J_{1}^{2}(kR(1-x^2)^{1/2})}{k^4 x^2(1-x^2)},
\end{equation}
where $R$ and $L$ are the comoving radius and height of the
cylinder,~$R= \Delta \theta (1+z) D_{\rm A}(z)/2$ and~$L\approx
(1+z)^2c (\Delta \nu /\nu_*)/H(z)$.

In equation~\eqref{obsvari},
$\beta(z)$ is the bias factor for the UCMH
clustering to the matter density fields. We assume that $\beta$ can be
obtained by the flux weighted average of the linear bias over the different collapse redshifts,
\begin{equation}
\beta(z)=\frac{\int^{z_{\rm{c,max}}}_{z_{\rm{c,min}}} b \Delta
\nu_{\rm{eff}}(z) \delta T_{{\rm b},z_{\rm c}}(z) {\cal S}
\frac{\mathrm{d}n}{\mathrm{d}z_{\rm c}}
\mathrm{d}z_{\rm c}}{\int^{z_{\rm{c,max}}}_{z_{\rm{c,min}}} \Delta \nu_{\rm{eff}}(z)
\delta T_{{\rm b},z_{\rm c}}(z) {\cal S} \frac{\mathrm{d}n}{\mathrm{d}z_{\rm c}} \mathrm{d}z_{\rm c}},
\end{equation}
where $b$ is the linear bias in the peak theory~(see equation~(24.b) with equations~(17) and (25) in~\citealt{1997MNRAS.284..189M}).
With the redshift decreasing, $b$ is monotonically becomes small.

\begin{figure}
    \includegraphics[width=\columnwidth]{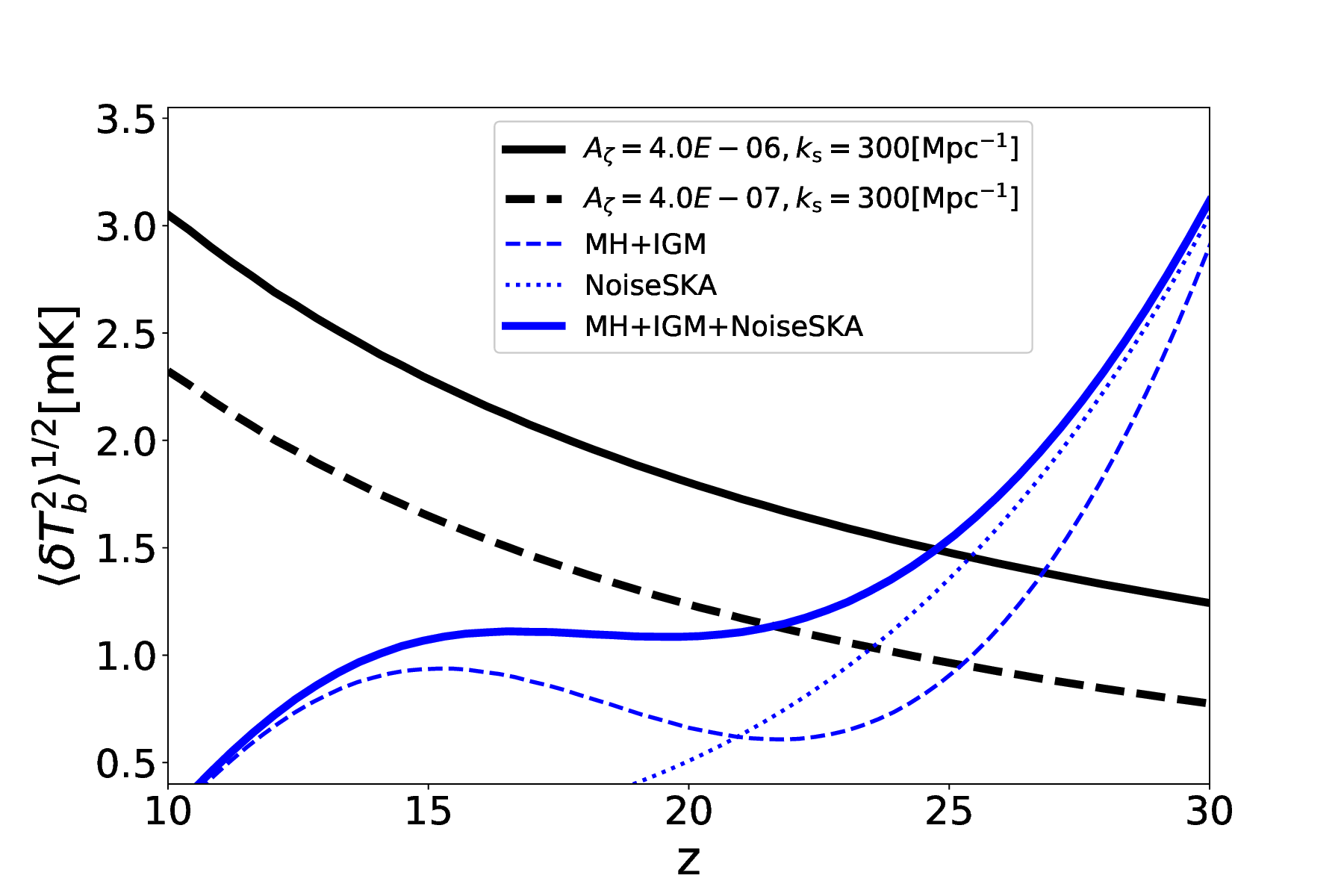}
 	\includegraphics[width=\columnwidth]{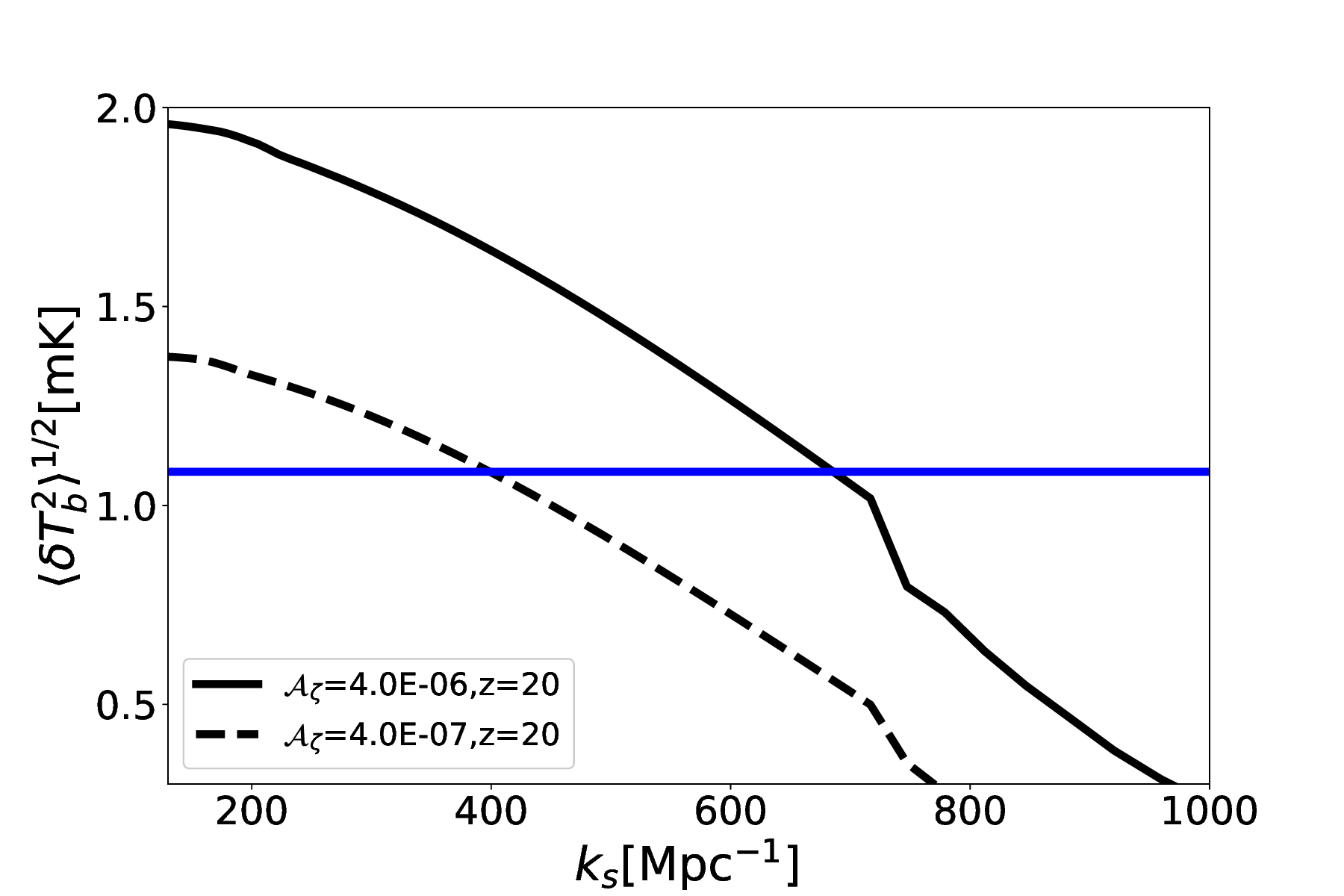}
     \caption{(\textit{Top panel}) Redshift dependence of the 21-cm fluctuations from UCMHs. The black solid line and the dashed lines are the 21-cm fluctuations from UCMH for $A_{\zeta}= 4.0 \times 10^{-6}$ and $4.0 \times 10^{-7}$, respectively. In both lines, we set $k_{\rm s}=300~{\rm Mpc^{-1}}$. We also plot the fluctuation from the IGM + MHs in the blue dotted line, the SKA noise level in the blue dot-dashed line and the sum of them in the blue solid line. (\textit{Bottom panel}) Dependence of the 21-cm fluctuations on the spike wave number $k_{\rm s}$ at $z=20$. The solid line is for $A_{\zeta}= 4.0 \times 10^{-6}$ while the dashed line is for $4.0 \times 10^{-7}$. The blue solid line represents the fluctuation from the IGM + MHs with the SKA noise level.}
     \label{graph6}
 \end{figure}

First we show the redshift evolution of the 21-cm fluctuations
of UCMHs in the top panel of figure~\ref{graph6}.
When the redshift decreases,
the fluctuations also become large
as same as 
the single UCMH signal in
figure~\ref{fig:signle_resdhift_evo}.
Note that we do not consider the impacts of cosmic reionization process. The abundant ionization photon background during the EoR can photoevaporate the neutral gas in UCMHs similarly to the case of minihaloes~\citep{2005MNRAS.361..405I}. 
This photoevapolation suppresses the amplitude of the fluctuations near the EoR, $z\lesssim 10$.

In the top panel of figure~\ref{graph6}, we also show the dependence of the fluctuations on the amplitude ${\cal A}_\zeta$.
The increment of ${\cal A}_\zeta$ enhances the fluctuations through two effects.
The first one is to increase the number density of the UCMHs.
However in our interested scales and redshifts, the number density is
saturated when $A_\zeta > 10^{-7}$. Hence, the enhancement through the
number density is weak.
The second one is that increasing ${\cal A}_\zeta$ results in the early-time
formation of UCMHs. As shown in figure~\ref{fig:signle_resdhift_evo},
the UCMHs forming in high redshifts provide large signals.
This effect provides
the dependence on ${\cal A}_\zeta$ in the top panel of figure~\ref{graph6}.

The bottom panel in figure~\ref{graph6} shows the dependence on $k_{\rm
s}$ for two different ${\cal A}_\zeta$ at $z=20$.
As $k_{\rm s}$ increases, the individual UCMH signal
becomes small. Following its signals, the rms fluctuations also decline. 
When $k_{\rm s}$ is larger than the Jeans wave number, the signal rapidly
drops down as expected from the section figure \ref{fig:single_ks}. 

\subsection{Detectability with SKA}

Several ongoing observations are attempting  
to detect the 21-cm fluctuation signal coming from $z \gtrsim 10$,~e.g.,
LOFAR, GMRT, MWA, and Paper.
Although there has been no report of the signal detection yet,
some observations have provided upper limits on the 21-cm fluctuations, e.g.,
$\delta T_{\rm b}^2 (k) <(49.0~{\rm mK})^2$
at $k\approx 0.59 h {\rm Mpc}^{-1}$ at
$z=6.5$~by MWA~\citep{2019ApJ...887..141L}
and
$\delta T_{\rm b}^2 (k) <(79.6~{\rm mK})^2$
at $k\approx 0.053 h {\rm Mpc}^{-1}$ at 
$z\sim10$ by LOFAR~\citep{2017ApJ...838...65P}.

However the maximum fluctuations by UCMHs in our model, $\delta T_{\rm b}^2  \lesssim(5~{\rm mK})^2$ at $z=20$, is fairly smaller than sensitivities of these instruments, we cannot provide the constraint on the UCMH abundance and the spike-shape power spectrum
by the current performance of the 21-cm fluctuation observations.

The upcoming 21-cm observation, SKA, is expected to detect and measure the 21-cm fluctuations at $z \gtrsim 10$.
We discuss the detectability of the 21-cm signals from UCMHs with the SKA.
In this paper, to discuss the detectability for the 21-cm signal from UCMHs, we estimate the 21-cm fluctuations by the IGM and minihaloes (MHs) and the observational noise of the SKA as well as the one by UCMHs.

In figure~\ref{graph6}, we plot the expected fluctuations due to the IGM + MHs and the noise of SKA in the blue lines. 
For the detailed calculation of them, we refer the reader to the Appendix.
We assume that the major contribution to the noise is the foreground emission of the sky. 
Since the foreground has the strong frequency
dependence, the noise rapidly blows up as the redshift increases~(the observation frequency decreases).
In the frequency and angular resolutions of our interest, the noise becomes larger than
the IGM + MH fluctuations above $z \sim 20$.
Note that we ignore the effect of cosmic reionization, which might enhance the 21-cm fluctuations of the IGM below $z \sim 10$ based on the Planck constraint on Thomson scattering optical depth.

\begin{figure}
 	\includegraphics[width=\columnwidth]{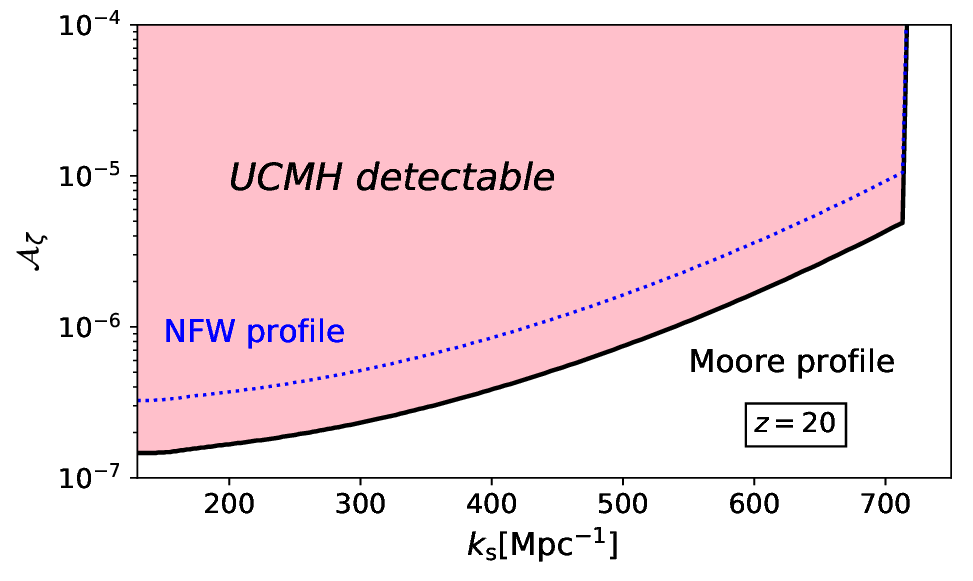}
     \caption{
      Detectability of the signal from UCMHs with observation by SKA. 
      The spike-shape parameter in the red region can produce
      the 21-cm fluctuations due to UCMHs which can be observed 
      by SKA. Here we set the observation redshift $z=20$.
      The wave number on the left end of the red region corresponds to $k_{\rm J}$. 
      The dotted line represents the lower bound for the detection,
      when a UCMH dark matter distribution takes the NFW profile.
 }
 \label{fig:const}
 \end{figure}

Our calculation has two free parameters related to the spike-shape spectrum,~$k_{\rm s}$ and ${\cal A}_\zeta$, and the amplitude of the fluctuations depends on them.
Taking the different parameter set,~($k_{\rm s}$, ${\cal A}_\zeta$), we survey the parameter region in which the 21-cm fluctuations of UCMHs can dominate the IGM + MH fluctuations and noise at $z=20$. We show such parameter region as the red region in figure~\ref{fig:const}. When there exists the fluctuations due to the spike-shape spectrum in the red region, the SKA can detect the excess of the 21-cm fluctuations from the expected IGM + MH signals at $z=20$. 
In other words, when the SKA does not find the excess of the 21-cm fluctuations, the delta-shape spectrum in the red region can be ruled out. 

At $k_{\rm s} \sim 700~{\rm Mpc}^{-1}$, one can see the sharp cutoff in figure~\ref{fig:const}.
The cutoff scale corresponds to the Jeans scale at the observation redshift~$z=20$.
When $k_{\rm s} > k_{\rm J}$, 
the baryon mass fraction radically drops down in UCMHs, in particular, which form late time. These UCMHs cannot host enough amount of the baryon gas to produce strong 21-cm signals for the detection.
As a result, 21-cm observations cannot provide the constraint below the Jeans scale.

Although we show the constraint only with the observation redshift~$z=20$, it is worth mentioning about its dependence on the observation redshift.
As shown in the top panel of figure \ref{graph6}, the fluctuation amplitude increases with the redshift decreases with fixing $k_{\rm s}$. Therefore, when we take a redshift lower than $z=20$, the constraint slightly becomes tighter.
Moreover as the redshift decreases, the Jeans scale becomes small. Accordingly, the sharp cutoff of the allowed region on the high-$k$ side shifts toward large~$k$. 
When we perform the 21-cm observation at~$z=10$, we can extend the red region to $k_{\rm s} \approx 1000~{\rm Mpc}^{-1}$~which corresponds to the Jeans scale at $z=10$.
However, in lower redshifts, the cosmological reionization process creates the additional 21-cm fluctuations as mentioned above. The additional fluctuations depend on the model of the reionization process. Therefore, it is difficult to evaluate the possible constraint on the primordial curvature amplitude with considering the reionization impact.

At the last of this section, we discuss the impact of the density profile on the 21-cm signals. 
In our calculation, we take the Moore profile and evaluate the gas density and temperature profiles based on the hydrostatic equilibrium.
However, according to numerical simulations by \citet{2017PhRvD..96l3519G}, if the UCMH formation occurs in non-isolated situations,
the profile of the resultant UCMHs
can be well fit by the NFW profile.
To investigate the profile dependence on the 21-cm signal, we simply adopt the NFW profile as the dark matter density distribution with using equations~\eqref{eq:rhos}~and~\eqref{eq:rs}, and evaluate the signal taking the procedure in section \ref{sec:theory} and \ref{flusec}.
Since the NFW profile has a shallower slope in the inner region than the Moore profile, the signal from the NFW halo is weaker than that from the Moore halo. 
We plot the critical amplitude for the detection in the blue dotted line in figure~\ref{fig:const}.
In short, when a UCMH has a shallower profile, the constraint on the spike-shape amplitude becomes loose.

\section{conclusion} \label{conc}

UCMHs are originated in the density fluctuations with the spike-shape spectrum on small scales. 
Therefore, the observational limit on the UCMH abundance leads the upper bound on the small-scale power spectrum of the primordial fluctuations. 
In this paper, we have evaluated the 21-cm signals from the UCMHs and have demonstrated that the SKA can provide the upper limit on the abundance of the UCMHs as well as the small-scale primordial fluctuations.

Since the 21-cm signal depends on the distribution of baryon gas in UCMHs, we have constructed the theoretical model of the baryon gas density and temperature, taking the Moore profile as the dark matter density distribution in UCMHs. 
Then we are first to have calculated the 21-cm signal from an individual UCMH. 
As a result, we have found that the individual signal becomes large as the UCMH mass increases.

Our observable are the 21-cm fluctuations caused by the number density fluctuations of UCMHs. 
We have investigated a relation between the 21-cm fluctuations and the spike-shape spectrum of the primordial fluctuations. 
We have shown that the 21-cm fluctuations by UCMHs become large as the amplitude of the spike-shape increases or the peak shifts to small scales. 
We also have found that the 21-cm fluctuations due to UCMHs are in an order of mK in the redshift range from $z=10$ and $30$. 
However, if the wave number of the spike-shape spectrum becomes smaller than the Jeans scale, the baryon gas mass in a UCMH drastically decreases, leading to the significant decrement of the signal. 

Though the 21-cm fluctuations by UCMHs does not reach the sensitivity of the current observational instruments, the SKA may detect the signal if there is the spike-shape spectrum we have discussed on the primordial power spectrum.

On the other hand, assuming non-detection of the signal by the SKA, an upper limit on the curvature perturbations is provided to 
${\cal A}_{\zeta} \lesssim 10^{-6}$ on $100~{\rm
Mpc}^{-1}\lesssim k \lesssim 1000~{\rm Mpc}^{-1}$.
Constraints on the UCMH abundance and the spike-shape spectrum have been provided through the gamma-ray observation assuming a self-annihilation of dark matter particles in UCMHs. 
Note that the constraint by the gamma-ray observation strongly relies on a model of dark matter, e.g., the annihilation mechanism and the mass of dark matter. In contrast, the 21-cm emission is independent of these detailed properties of dark matter. 
Therefore, the probe of 21-cm fluctuations by future observations is expected to be a more robust survey of the primordial fluctuation on small scales.

\section*{Acknowledgements}
Numerical calculations were partially carried out on
clusters installed in Nagoya University. 
This work was supported by JSPS KAKENHi Grant Numbers JP17H01110 (KH, HT, TT) and JP18K03699 (KH).

\appendix

\section{21-cm signals from IGM and MHs} \label{consig}

In this appendix, we give descriptions for construction of 21-cm signals from IGM and MHs.
Here, we assume $x_{\alpha}=0$ in Eq. (\ref{ts}) because we assume that no UV and X-ray sources exist in the Universe above $z=20$.

\subsubsection*{IGM fluctuations}

The mean differential brightness temperature of the IGM is given by
~\citep{1997ApJ...475..429M}
\begin{equation}
\overline{ \delta T_{\rm b}} = 9.1
 \left({1+z} \right)^{1/2} \left(1-\frac{T_{\rm{CMB}}}{T_{\rm{S}}}\right)
\left( \frac{\Omega_{\rm b} h}{0.33} \right)
  \left(\frac{\Omega_{\rm m}}{0.27} \right)^{-1/2}~{\rm mK},
\end{equation}
where we assume that the IGM is fully neutral.
For the IGM spin temperature, we take the IGM gas temperature evolution
with only the Compton heating.

The 21-cm signals spatially fluctuate, depending on the baryon density
fluctuations. We assume that the baryon distribution follows the matter
density fluctuations.
Accordingly, the observed rms 21-cm fluctuations due to the IGM can
be evaluated in
\begin{equation}
\langle \delta T_{\rm{b,IGM}}^2 \rangle^{1/2} = 
\sigma_{p}(z) \overline{ \delta T_{\rm b}},
\end{equation}
where $\sigma_{p}(z)$ is given in equation~\eqref{obsvari}.

In our evaluation, we do not take into account the
contribution of the ionization fluctuation in the IGM.
This contribution to the 21-cm fluctuations increases
as the cosmological reionization proceeds.
Therefore, our evaluation on the IGM fluctuations is
underestimated around $z\sim 10$.

\subsubsection*{Minihalo fluctuations}

minihaloes~(MHs) are virialized objects with 
the virial temperature $T_{\rm{vir}} < 10^4~$K.
In the standard hierarchical structure formation,
they form abundantly in high redshifts
and are filled with neutral hydrogen atoms.
Therefore MHs are promised sources of 21-cm signals in high redshifts~\citep{2002ApJ...572L.123I, 2006ApJ...652..849F}.

In figure~\ref{graph6}, we evaluate the 21-cm fluctuations due to MHs in the same manner as in the case of UCMHs, except for the dark matter profile.
It is predicted that the dark matter distribution in MHs can be described in the NFW profile.
For the concentration parameter, $p_{\rm con} \equiv {r_{\rm{vir}}}/{r_{\rm{s}}}$ in the NFW profile,
we set~\citep{2000MNRAS.318..203S}
\begin{equation}
p_{\rm con}=\frac{10}{1+z_{\rm c}} \left[\frac{M}{M_{*}(z=0)}\right]^{-0.2},
\end{equation}
Here $M_{*}(z)$ satisfies $\sigma(M_*,z) = \delta_{\rm{c}}$ where $\sigma(M,z)$ is the dispersion of the density fluctuations smoothed with the top-hat filter of the radius corresponding to mass $M$ at the redshift~$z$.
In the calculation, we assume that
the mass range of MHs spans from
the Jeans mass $M_{\rm{J}}$ to the virial mass with $T_{\rm{vir}}=10^4~$K.

\subsubsection*{Noise level of SKA}

The noise level of an interferometer such as SKA is
given with the observation wave length $\lambda$ by~\citet{2006PhR...433..181F}
\begin{equation}
\delta T_{\rm{N}}(\lambda)=
\frac{\lambda^2}{A_{\rm eff} \Delta \theta^2}
\frac{T_{\rm sys}}{\sqrt{ \Delta \nu t_{\rm obs}}},
\end{equation}
where $A_{\rm{eff}}$, $t_{\rm{obs}}$ and
$T_{\rm sys}$are the effective collecting area, the observational time,
and the system temperature, respectively.

We are interested in low frequency observation, $\nu <150~\rm Hz$.
In such low frequency region,
the sky temperature is the dominant component om $T_{\rm sys}$.
We take the sky temperature at high Galactic latitude where the
foreground is low in the sky,
\begin{equation}
 T_{\rm sys} \sim 180 \left(\frac{180~{\rm MHz}}{\nu}\right)^{2.6} {\rm K}.
\end{equation}

Therefore, the noise level is given in
\begin{eqnarray}
\delta T_{\rm{N}}(\nu) = 0.507 \rm{m K} \left(\frac{8 \times 10^5 \rm{m}^2}{A_{\rm{eff}}}\right)
\left(\frac{20'}{\Delta \theta}\right)^{2}
\left(\frac{3\rm{MHz}}{\Delta \nu}\right)^{1/2}
\nonumber \\ \times
\left(\frac{1000 \rm{h}}{t_{\rm{obs}}}\right)^{1/2}
\left(\frac{1+z}{21} \right)^{4.6}.
\end{eqnarray}
For SKA, we set 
$A_{\rm{eff}}=8 \times 10^5 ~\rm{m}^2$, $\Delta \theta=20~\rm{arcmin}$,  $\Delta \nu=3~\rm{MHz}$ and $t_{\rm{obs}}=1000~\rm{h}$.

\bibliographystyle{mnras}
\bibliography{bibdata}

\end{document}